\documentclass[twocolumn,letterpaper,showpacs]{revtex4}

\usepackage{graphicx}
\usepackage{dcolumn}
\usepackage{amsmath}
\usepackage{amssymb}
\usepackage{float}

\begin{document}

\title{Exploring Spiral Defect Chaos in Generalized Swift-Hohenberg Models with Mean Flow}
\author{A. Karimi}
\affiliation{Department of Engineering Science and Mechanics, Virginia
Polytechnic Institute and State University, Blacksburg, Virginia 24061}
\author{Zhi-Feng Huang}
\affiliation{Department of Physics and Astronomy, Wayne State University, Detroit, MI 48201}
\author{M. R. Paul}
\email{mrp@vt.edu}
\affiliation{Department of Mechanical Engineering, Virginia
Polytechnic Institute and State University, Blacksburg, Virginia 24061}


\begin{abstract}

We explore the phenomenon of spiral defect chaos in two types of
generalized Swift-Hohenberg model equations that 
include the effects of long-range drift velocity or mean flow. We use 
spatially-extended domains and integrate the equations for very long times to study 
the pattern dynamics as the magnitude of the mean flow is varied.  The magnitude 
of the mean flow is adjusted via a real and continuous parameter that accounts 
for the fluid boundary conditions on the horizontal surfaces in a convecting layer. For 
weak values of the mean flow we find that the patterns exhibit a slow coarsening to a 
state dominated by large and very slowly moving target defects. For
strong enough mean flow we identify the existence of 
spatiotemporal chaos which is indicated by a positive leading order Lyapunov 
exponent. We compare the spatial features of the mean flow field with that of 
Rayleigh-B\'enard convection and quantify their differences in the neighborhood of 
spiral defects. 
\end{abstract}


\maketitle

\section{Introduction}

The chaotic behavior of spatially-extended dissipative systems has been 
intensively studied in recent years~\cite{cross:1993}. Spatiotemporal chaos 
has been observed in a wide range of physical
systems including Faraday waves that appear on the surface of an 
oscillating layer of fluid~\cite{gollub:1991}, reacting chemical 
mixtures~\cite{skinner:1991}, excitable media~\cite{bar:1993}, 
and Rayleigh-B\'{e}nard convection in a shallow fluid layer heated 
from below~\cite{cross:1993}. In particular, the study of Rayleigh-B\'{e}nard 
convection continues to provide fundamental insights into the 
dynamics of pattern forming systems that are driven 
far-from-equilibrium~\cite{cross:1993,bodenschatz:2000}. The state of spiral 
defect chaos has received significant attention
since its discovery by Morris \textit{et al.}~\cite{morris:1993}.
Spiral defect chaos is characterized by the complex dynamics of
rotating spiral defects and interestingly, occurs for fluid parameters
where straight parallel convection rolls are linearly 
stable~\cite{cross:1993,bodenschatz:2000}. Although much effort has been 
spent on building our understanding of the origins and dynamical 
features of spiral defect chaos, many open questions still remain.

A significant difficulty in studying spatiotemporal chaos is that the experimental 
systems of interest are often large and strongly driven. 
This presents significant obstacles for both analytical and numerical approaches. 
For example, the numerical simulation of the Boussinesq equations, which govern 
Rayleigh-B\'enard convection, in large domains, for long simulation times, and 
for many values of the system parameters is out of reach using currently available algorithms 
and computing resources. Although significant progress has been made in the 
ability to simulate convection for experimental conditions, the computational 
cost remains very high~\cite{paul:2003}. In light of challenges such as these, the 
use of simpler model equations has played a pivotal role in furthering our physical 
understanding of spatiotemporal chaos.  In Rayleigh-B\'{e}nard convection, the 
two-dimensional Swift-Hohenberg equation~\cite{swift:1977} has led to 
numerous physical insights regarding questions of pattern formation~\cite{cross:1993}. 
However, the use of the Swift-Hohenberg equation to study spatiotemporal chaos, and 
in particular spiral defect chaos~\cite{xi:1993}, has been called into question~\cite{schmitz:2002},
as will be further discussed below.  This leaves no clear choice for a model
system to be used for the study of spiral defect chaos with direct relevance to 
fluid convection.

The presence of a long-range mean flow is well known to play an 
important role in the dynamics of Rayleigh-B\'enard convection
\cite{cross:1993,newell:1990:jfm,newell:1990:prl}. 
This led to extensions of the Swift-Hohenberg equation to account for these 
effects~\cite{manneville:1984,cross:1995:prl,xi:1993}. The mean flow is a 
weak horizontal flow field that acts over length scales larger than that of 
the convection rolls; it results from the coupling to fluid vertical
vorticity and is induced by roll curvature, amplitude 
gradients, and wavenumber gradients~\cite{re:siggia81,manneville:1983,cross:1993}. 
The magnitude of the mean flow is much smaller than that of the convective roll 
motion, making it very difficult to measure in experiment.  A typical
way to visualize the mean flow
in numerical work is to present contours of the integrated horizontal components 
of the fluid velocity over the depth of the fluid layer~\cite{paul:2003}.  Although the 
mean flow is weak it can have significant effect upon the dynamics and stability 
boundaries of the flow field and also adds a slow time scale to the dynamics.  
It has been shown numerically that the mean flow is required for spiral defect 
chaos in Rayleigh-B\'{e}nard convection~\cite{chiam:2003}.

Spiral defect chaos was initially explored using numerical simulations of the 
generalized Swift-Hohenberg equation by Xi \textit{et al.}~\cite{xi:1993}.  
These simulations were for rather short intervals of time $t \sim 900$ 
where $t$ is the nondimensional time. We note that to generate the
spiral defect state in this model system, the system needs to be taken
far enough from the convective threshold; otherwise roll-type patterns 
would dominate with the system dynamics governed by the motion of topological
defects such as grain boundaries, dislocations, and disclinations
\cite{re:huang07}. Schmitz \emph{et al.}~\cite{schmitz:2002} 
later explored the generalized Swift-Hohenberg equation for the same parameters 
but for much longer simulation times $t \sim 64000$. Their results suggested 
that the spiral defect state was only a transient with the long-time dynamics 
characterized by a slow coarsening process to a state dominated by large spirals. 
This work has cast doubt upon the ability of the Swift-Hohenberg 
equation to exhibit persistent dynamics that resemble spiral defect chaos.

In this paper, we present a careful numerical study of these questions. In 
particular, we have performed very long-time simulations ($t = 10^6$ time units) 
in very large domains, and for a wide range of mean flow strengths.  We also 
compute the leading order Lyapunov exponent to determine if the dynamics 
are chaotic. Our investigation is driven by the results on Rayleigh-B\'{e}nard 
convection that demonstrate the importance of mean flow upon spiral
defect chaos~\cite{paul:2003,bodenschatz:2000}.

The remainder of this paper is organized as follows. In
Sec.~\ref{section:approach} we discuss two types of generalized
Swift-Hohenberg model equations and provide some details regarding
its numerical solution and the computation of the leading order Lyapunov exponent. 
In Sec.~\ref{section:results} we present our results and discuss the role of the 
mean flow upon the dynamics.  Concluding remarks are given in
Sec.~\ref{section:conclusion}. 

\section{Approach}
\label{section:approach}

\subsection{The Generalized Swift-Hohenberg Models}
\label{section:models}

A generalized, dimensionless form of the two-dimensional time-dependent
Swift-Hohenberg model is given by~\cite{xi:1993}
\begin {eqnarray}
	\frac{\partial \psi}{\partial t}+\mathbf{U} \cdot \nabla \psi=\epsilon \psi - (\nabla^2+1)^2 \psi+ N[\psi],  \label{eq:psi}\\
	\left[\frac{\partial}{\partial t}-\sigma \left(\nabla^2-c^2 \right) \right] \nabla^2 \zeta = g_m \left[\nabla \left(\nabla^2 \psi \right)\times \nabla \psi \right]\cdot \hat{\mathbf{z}}, \label{eq:zeta}
\end {eqnarray}
where $\psi(x,y,t)$ is a scalar field describing the spatial and
temporal variation of the convection 
patterns, $\epsilon$ is a control parameter giving the dimensionless
distance from the convective
threshold, and $\hat{\mathbf{z}}$ represents the unit vector in the
out-of-plane direction. The variable $\zeta(x,y,t)$ is  
the vertical vorticity potential defined via $\Omega_z = (\nabla \times
\mathbf{U})_z = -\nabla^2 \zeta$,
where $\Omega_z$ is the vertical component of fluid vorticity and
$\mathbf{U}$ is the mean flow or drift velocity.
$\zeta$ can also be interpreted as the stream function for the mean
flow, given that
\begin {equation}
	\mathbf{U}=\nabla \times \left(\zeta \hat{\mathbf{z}} \right)=\left( \partial_y \zeta, -\partial_x \zeta \right). 
	\label{eq:U}
\end {equation}
In Eq. (\ref{eq:zeta}) $\sigma$ is proportional to the Prandtl number,
$g_m$ is a positive real constant that characterizes the mean flow
coupling strength, and $c$ is a real parameter introduced for modeling
the effect of free-slip ($c=0$) or no-slip ($c \neq 0$) boundary
conditions for the horizontal surfaces of a convection layer.  The
term $N[\psi]$ in Eq. (\ref{eq:psi}) represents the 
nonlinearity of the system, which has many different forms in the
literature based on system conditions
\cite{greenside:1985,cross:1993,manneville:1984}.
In this paper we present numerical results for two representative cases: 
(1) $N[\psi]=-\psi^3$ for which the model equations are referred to as
the Generalized Swift-Hohenberg (GSH) model~\cite{greenside:1985}, and 
(2) $N[\psi]= -( |\nabla \psi|^2 \psi + \psi^3 )$ for which the
equations are referred to as Manneville's model that was derived in
Refs.~\cite{manneville:1983,manneville:1984} from the Boussinesq equations.

\subsection{Numerical Simulations}
\label{section:simulations}

We numerically integrate Eqs.~(\ref{eq:psi})--(\ref{eq:zeta}) using the approach discussed by 
Cross \textit{et al.}~\cite{cross:1994} and provide only the essential details here.  The domain 
is a square geometry that is discretized on a spatially uniform grid with periodic lateral 
boundary conditions.  In our numerical simulations we begin from random initial conditions 
for $\psi$ and set initially $\zeta(x,y,t=0)=0$.  We discretize the spatial domain using 
a $512 \times 512$ grid with a grid spacing of $\Delta x=\Delta y=\lambda_0/8$ where 
$\lambda_0= 2 \pi$ is the critical wavelength of a convection roll. These parameters are 
approximately equivalent to a Rayleigh-B\'enard system in a box geometry with an aspect 
ratio of $\Gamma = L/d = 128$ where $L$ is the side length of the box and $d$ is the 
depth of the convection layer. Each individual simulation is allowed to evolve for 
$t=10^6$ time units using a time step of $\Delta t=0.2$.

The equations are stiff due to the very fast dynamics of the biharmonic operator in 
comparison to the much slower convective time scale of the pattern dynamics. An 
efficient solution is obtained using a pseudospectral
operator-splitting approach~\cite{canuto:1988}.
The linear terms are treated exactly using an explicit exponential time 
integration~\cite{cox:2002,kassam:2005} and the nonlinear terms are evolved forward 
in time using an explicit predictor-corrector approach. To reduce the contributions of 
high wavenumber modes in the vorticity field a Gaussian filtering
operator $F_\gamma$ is applied to the 
right-hand side of Eq.~(\ref{eq:zeta}) \cite{greenside:1985}.  In
Fourier space it is given by $F_{\gamma}=\exp \left(-{\gamma}^2 q^2/2
\right)$ where $\gamma$ is the filtering radius and $q$ is the
wavenumber.  In our simulations we have used a filtering radius of $\gamma=\lambda_0/2$. 

We compute the leading order Lyapunov exponent using the standard procedure described in 
detail in Ref.~\cite{wolf:1985}. The tangent space equations are,
\begin {eqnarray}
	\frac{\partial \delta \psi}{\partial t}+\mathbf{U} \cdot \nabla \delta \psi+\delta \mathbf{U} \cdot \nabla \psi=\epsilon \delta \psi - (\nabla^2+1)^2 \delta \psi \nonumber \\
	+\delta N[\psi, \delta \psi],  \label{eq:dpsi} \\
	\left[\frac{\partial}{\partial t}-\sigma \left(\nabla^2-c^2 \right) \right] \nabla^2 \delta \zeta = g_m \left[\nabla \left(\nabla^2 \delta \psi \right)\times \nabla \psi \right. \nonumber \\
	+ \left. \nabla \left( \nabla^2 \psi \right) \times \nabla \delta \psi \right]\cdot \hat{\mathbf{z}}, \label{eq:dzeta}
\end {eqnarray}
where $\delta \mathbf{U}=\nabla \times  \left(\delta \zeta
\hat{\mathbf{z}} \right)$. The nonlinear term for the
GSH equation is
\begin{equation}
	\delta N[\psi, \delta \psi]=-3 \psi^2 \delta \psi,
\end{equation}
and for Manneville's model it is 
\begin{equation}
	\delta N[\psi, \delta \psi]= - \left( 3 \psi^2 + | \nabla \psi |^2 \right) \delta \psi -2 \left[ (\nabla \psi) \cdot (\nabla \delta \psi) \right] \psi.
\end{equation}
The magnitude of $\delta \psi$ is renormalized after a time 
$t_N$ to yield a measure of its growth $\| \delta \psi(t_N) \|$ 
which is used to calculate the instantaneous Lyapunov exponent, i.e.,
\begin{equation}
	\tilde{\lambda}_1 = \frac{1}{t_N} \ln \| \delta \psi(t_N) \|.
\end{equation}
We have used $t_N=2$ in our simulations.  This normalization is repeated in 
time to generate many values of the instantaneous Lyapunov 
exponent whose time average yields the finite-time Lyapunov exponent
\begin{equation}
	\lambda_1 = \frac{1}{N_t} \sum_{i=1}^{N_t} \tilde{\lambda}_1
	\label{eq:lyapunov}
\end{equation}
where $N_t$ is the number of renormalizations performed. The limit
$N_t \rightarrow \infty$ yields the infinite-time Lyapunov exponent.

\section{Results}
\label{section:results}

Experimental measurements of spiral defect chaos have typically been performed using 
large aspect ratio domains, moderate Rayleigh numbers, and compressed gases with 
Prandtl number of approximately unity~\cite{bodenschatz:2000}.  Generalizations of 
the Swift-Hohenberg model, as given by Eqs.~(\ref{eq:psi})--(\ref{eq:zeta}), 
have been used to study fundamental features of spiral defect chaos. The choice 
of the system parameters in the Swift-Hohenberg-type models are important in order to 
yield dynamics that resemble spiral defect chaos.  In order to estimate the 
appropriate parameters, Xi~\textit{et al.}~\cite{xi:1993} compared a 
three-mode amplitude equation for the GSH equation 
with the experimental results of Ref.~\cite{bodenschatz:1991}.   Using this approach 
yielded values of $\sigma=1$, $\epsilon=0.7$, and $g_m=50$. It was also chosen 
to use $c^2=2$ although a specific physical reason for this particular choice 
is not given. This parameter set has been adopted in most numerical
explorations of spiral defect chaos using the Swift-Hohenberg model,
including that of Schmitz~\textit{et al.}~\cite{schmitz:2002} where it 
was suggested that spiral defect chaos in the numerical simulations 
was only a transient.

The choice of these system parameters highly affects the magnitude and
dynamics of the mean flow.  The magnitude of the mean flow is inversely 
proportional to $\sigma$ and increases with increasing values of the coupling strength 
$g_m$. Cross~\cite{cross:1996} has presented a careful study of the variation of 
the dynamics with $g_m$.  For small values of $g_m$ the patterns 
were dominated by target defects and for large values of $g_m$ the patterns 
were dominated by spirals. Schmitz~\textit{et al.}~\cite{schmitz:2002} computed 
the appropriate value of $g_m$ based upon the zig-zag stability boundary for 
$\epsilon=0.7$ and $\sigma=1$ and found that $g_m \approx 12$. However it was 
determined by numerical exploration that a larger value was required to yield 
dynamics that resembled spiral defect chaos and a value of $g_m=50$
was used in their numerics.

Our calculations here indicate an important role played by the parameter
$c^2$ on the strength of mean flows. As described above in
Sec. \ref{section:models}, $c^2$ is related to the choice of boundary
conditions on the bottom and top plates of a three-dimensional convection 
system.  Its value physically accounts for the viscous damping that occurs 
near the horizontal surfaces. A value of $c^2=0$ corresponds to perfect slip 
on the top and bottom plates and a value of $c^2 \ne 0$ corresponds 
to a no-slip boundary condition.  In the development of 
the GSH equation~\cite{greenside:1985} the term $c^2$ is 
introduced as an unknown constant.  In the derivation by Manneville 
~\cite{manneville:1983,manneville:1984} $c^2$ emerges as part of the
expansion and averaging procedures used when starting from the
Boussinesq equations.  Using the nondimensional form of the equations
shown in Eq.~(\ref{eq:psi})--(\ref{eq:zeta}) yields
a value of $c^2=1.03$ (after rescaling) for no-slip
boundaries~\cite{manneville:1984}. However, as pointed out in
Ref.~\cite{manneville:1984}, the precise numerical value of $c^2$
depends upon the approximation process as well as the manner in which the 
averaging is done in the vertical direction.

In the following we have explored the details of mean flow and spiral
defect state using both Manneville's model and the GSH equation.  What separates 
our work from previous efforts is that we have explored the role of the mean 
flow by systematically varying the value of $c^2$ while also computing the 
leading order Lyapunov exponent for very long-time simulations. When using 
Manneville's model we have chosen the system parameters to most 
closely align with those of a Rayleigh-B\'enard convection domain exhibiting 
spiral defect chaos (i.e., $\epsilon=0.7,\sigma=2,g_m=50$).  When using the 
GSH equation we have used the values commonly 
used in the literature (i.e., $\epsilon=0.7,\sigma=1,g_m=50$). For both models 
we have explored the range of $c^2$ values from $0.1 \le c^2 \le 4$.  We note that 
for our choice of the system parameters the Prandtl number is different between 
our simulations using Manneville's model and the GSH 
model. Our intention is not to provide a quantitative comparison between these 
two models but to explore the role of mean flow upon spiral defect chaos.

\begin{figure}[htb]
\begin{center}
	\includegraphics[width=3in]{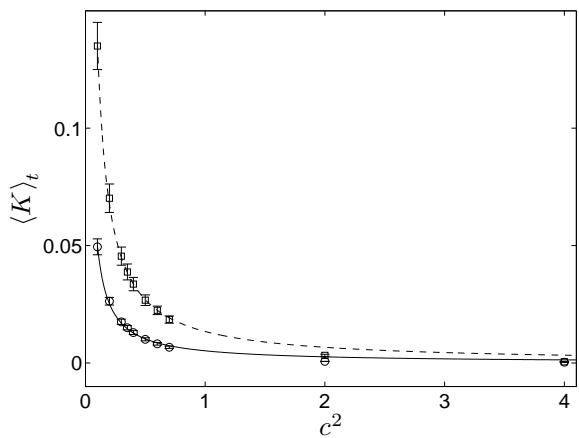}  
\end{center}
\caption{The variation of the spatial and time averaged kinetic energy $\left< K \right>_t$ 
with $c^2$ . The circle symbols are for Manneville's model with parameters
$\epsilon=0.7$, $\sigma=2$, $g_m=50$, and the square symbols are for the 
GSH equation with $\epsilon=0.7$, $\sigma=1$, $g_m=50$.
The lines are curve fits given by $\left< K \right>_t = 0.013 c^{-2}$
for the GSH model and $\left< K \right>_t = 0.005 c^{-2}$ for Manneville's model. 
The results are averaged using the last $10^5$ time units of a
simulation with total time 
$t=10^6$. The error bars are the standard deviation of the fluctuations 
of $\left< K \right>_t$ about its mean value; their maximum values are $\sim 10^{-2}$
and $\sim 10^{-3}$ for the GSH model and Manneville's model, respectively.} 
\label{fig:kinetic-energy}
\end{figure}

The variation of the mean flow magnitude with $c^2$ can be quantified 
by computing the average kinetic energy of the mean flow field. The time dependent 
value of the spatially averaged kinetic energy is given by 
\begin {equation}
	K(t) =\frac{1}{2} \left< \left( U_x^2+U_y^2 \right) \right>.
	\label{eq:kinetic-energy}
\end{equation}
where ($U_x,U_y$) are the $x$ and $y$ components of the mean flow 
velocity and $\left< ... \right>$ represents the spatial average. 
In Fig.~\ref{fig:kinetic-energy} 
we plot $\left< K \right>_t$ which represents the time averaged value
of $K(t)$ as a function of $c^2$. The time averaging is performed from the data 
of the final $10^5$ time units. The error bars represent the standard
deviation of the variation of $\left< K \right>_t$
about its mean value. The circle symbols are numerical results for Manneville's 
model and the square symbols are results for the GSH equation.  
For both models the data can be fitted to $\left< K \right>_t \propto c^{-2}$, 
indicating the rapid increase in the magnitude of the mean flow for decreasing 
values of $c^2$.

\begin{figure}[htb]
\begin{tabular}{cc}
	   \includegraphics[width=1.5in]{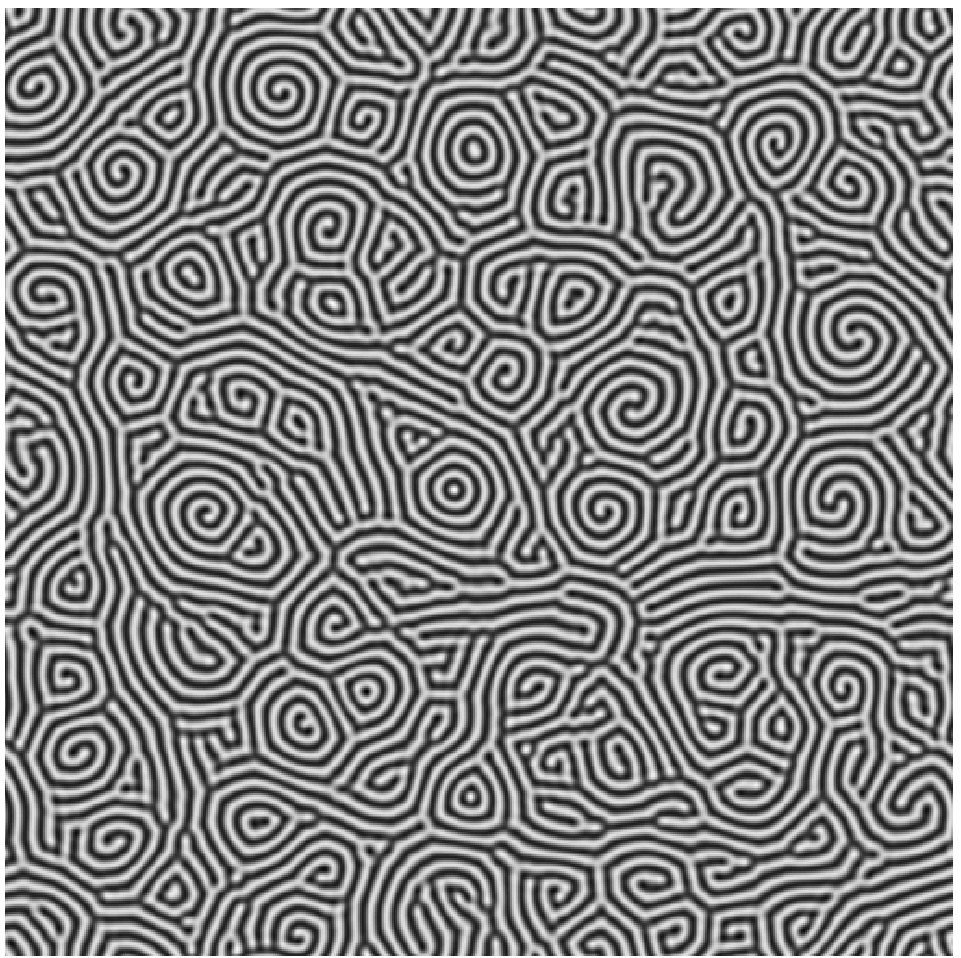} \hspace{0.2cm} &
	   \includegraphics[width=1.5in]{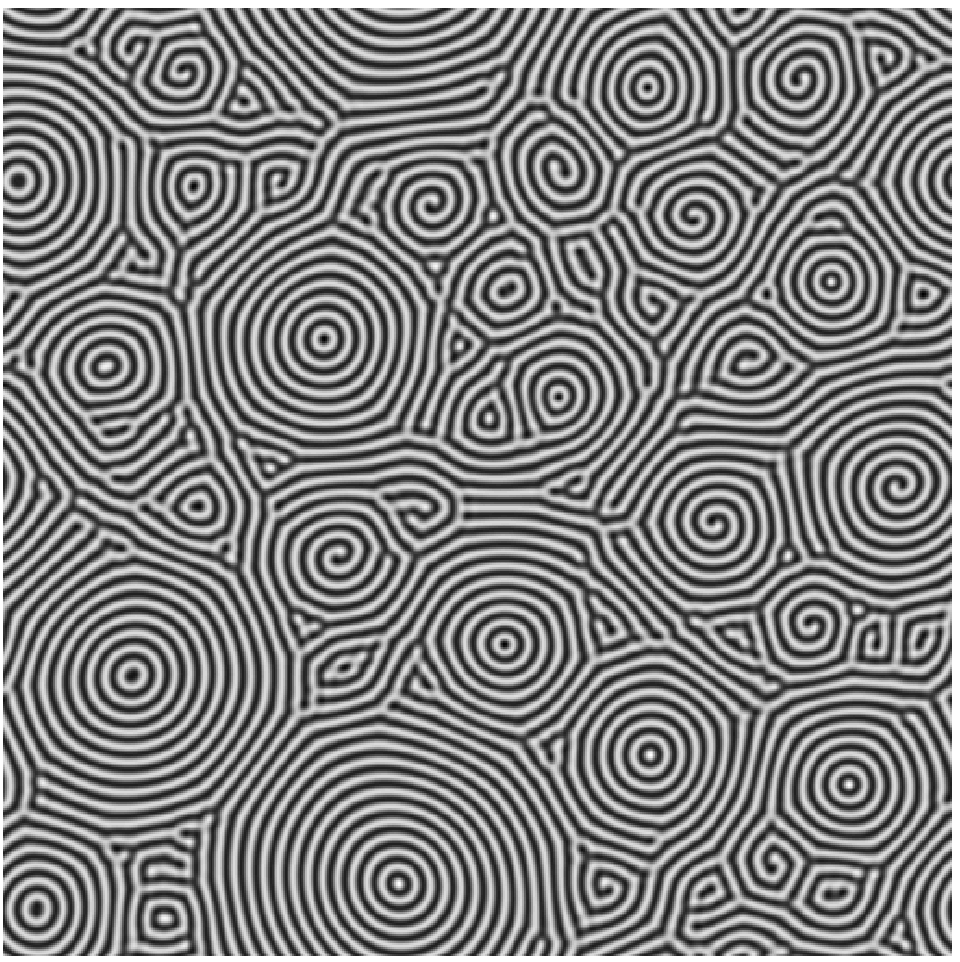} \\
	   $t=1000$  &  $t=50,000$ \vspace{0.2cm} \\
	   \includegraphics[width=1.5in]{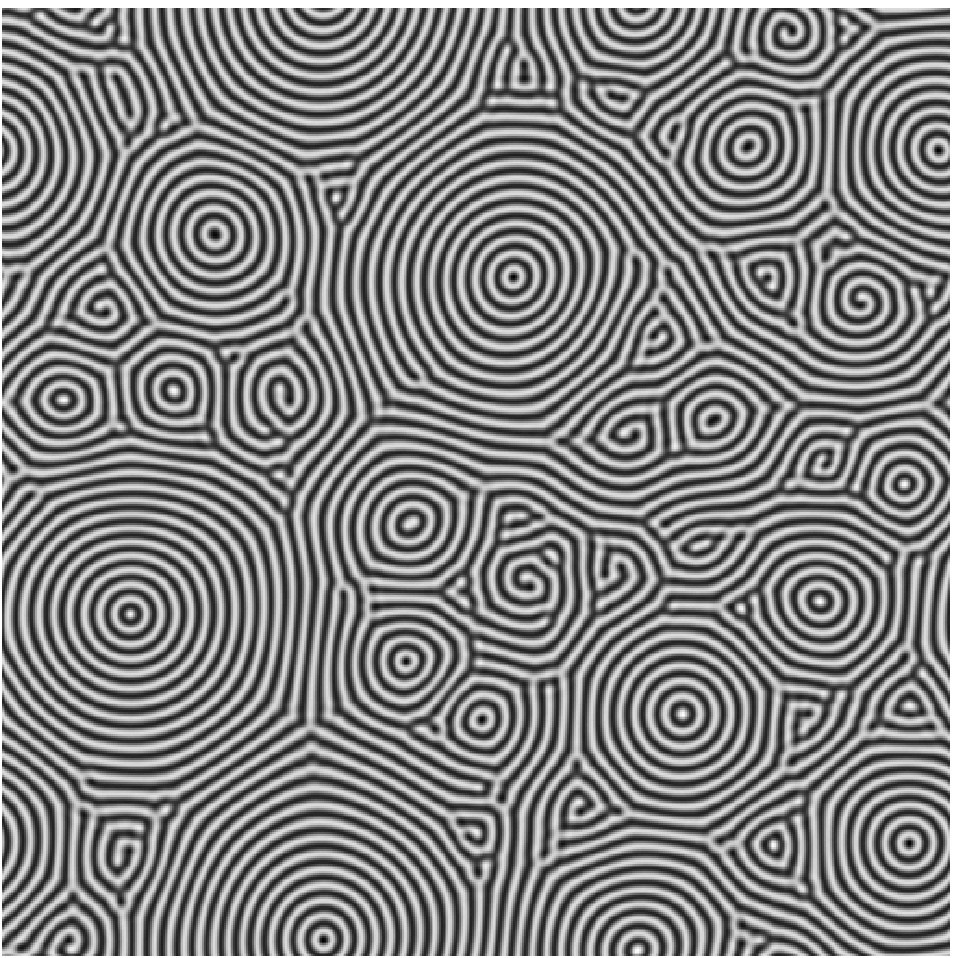} \hspace{0.2cm} &
	   \includegraphics[width=1.5in]{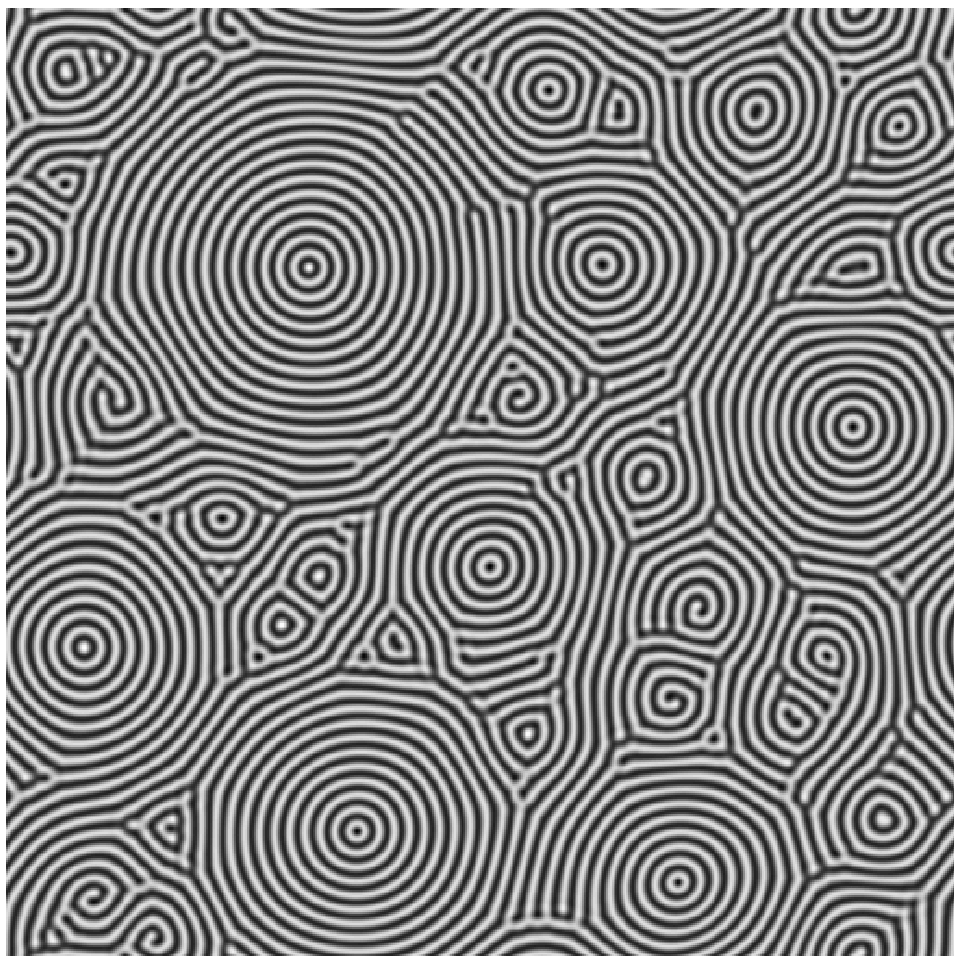} \\
	   $t=100,000$  &  $t=200,000$ \\
\end{tabular} 
\caption{Instances of the $\psi$ field pattern for Manneville's model with 
system parameters $\epsilon=0.7$, $\sigma=2$, $g_m=50$, and $c^2=2$,
showing a slow process of pattern coarsening. Small spiral defects
dominate at early times (e.g., $t = 10^3$), whereas at large 
times the pattern has evolved into a state 
dominated by large and very slow moving target structures.} 
\label{fig:coarsening}
\end{figure}

\begin{figure}[htb]
\begin{tabular}{cc}
	\includegraphics[width=1.5in]{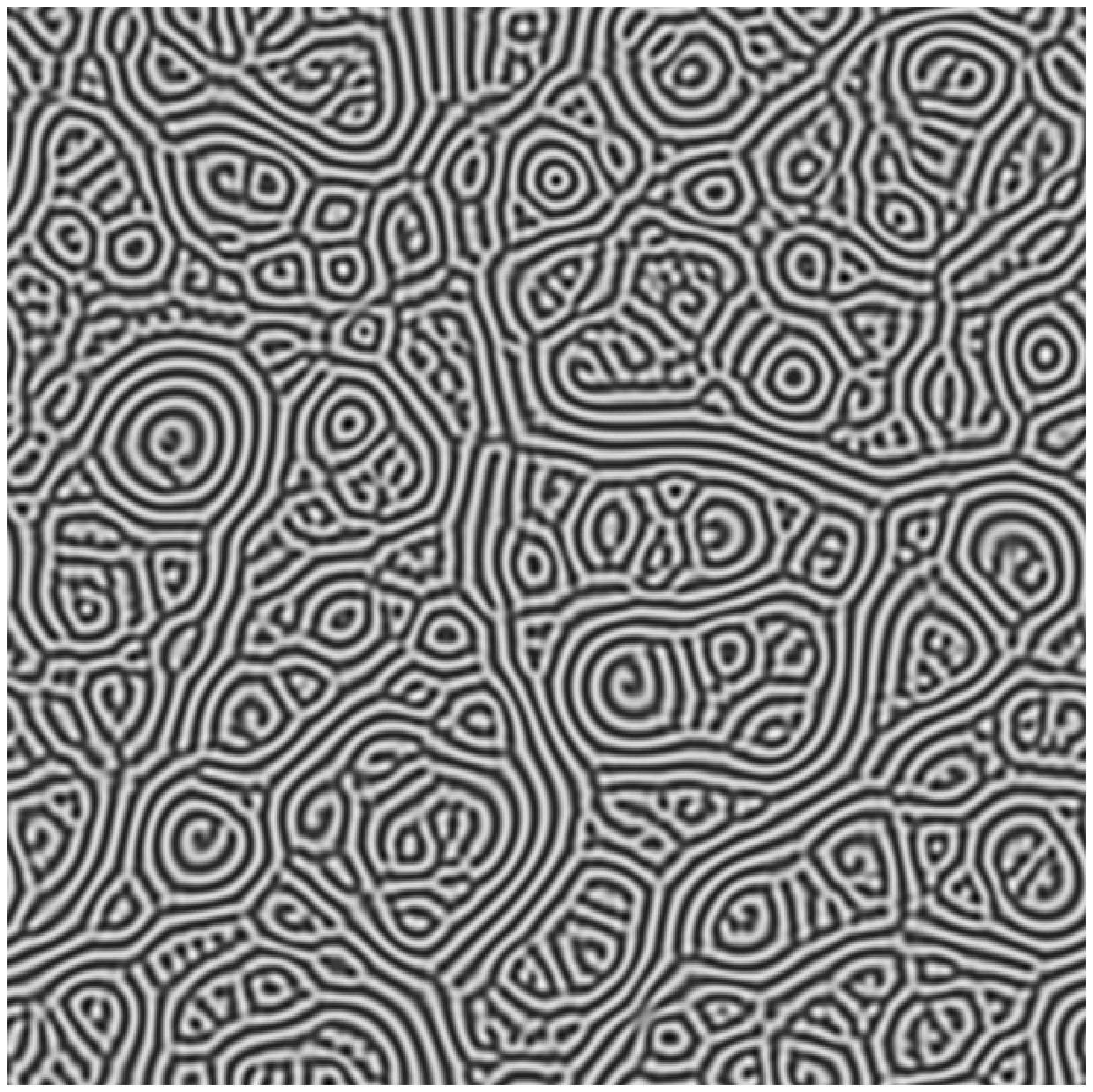}  \hspace{0.5cm} & 
	\includegraphics[width=1.5in]{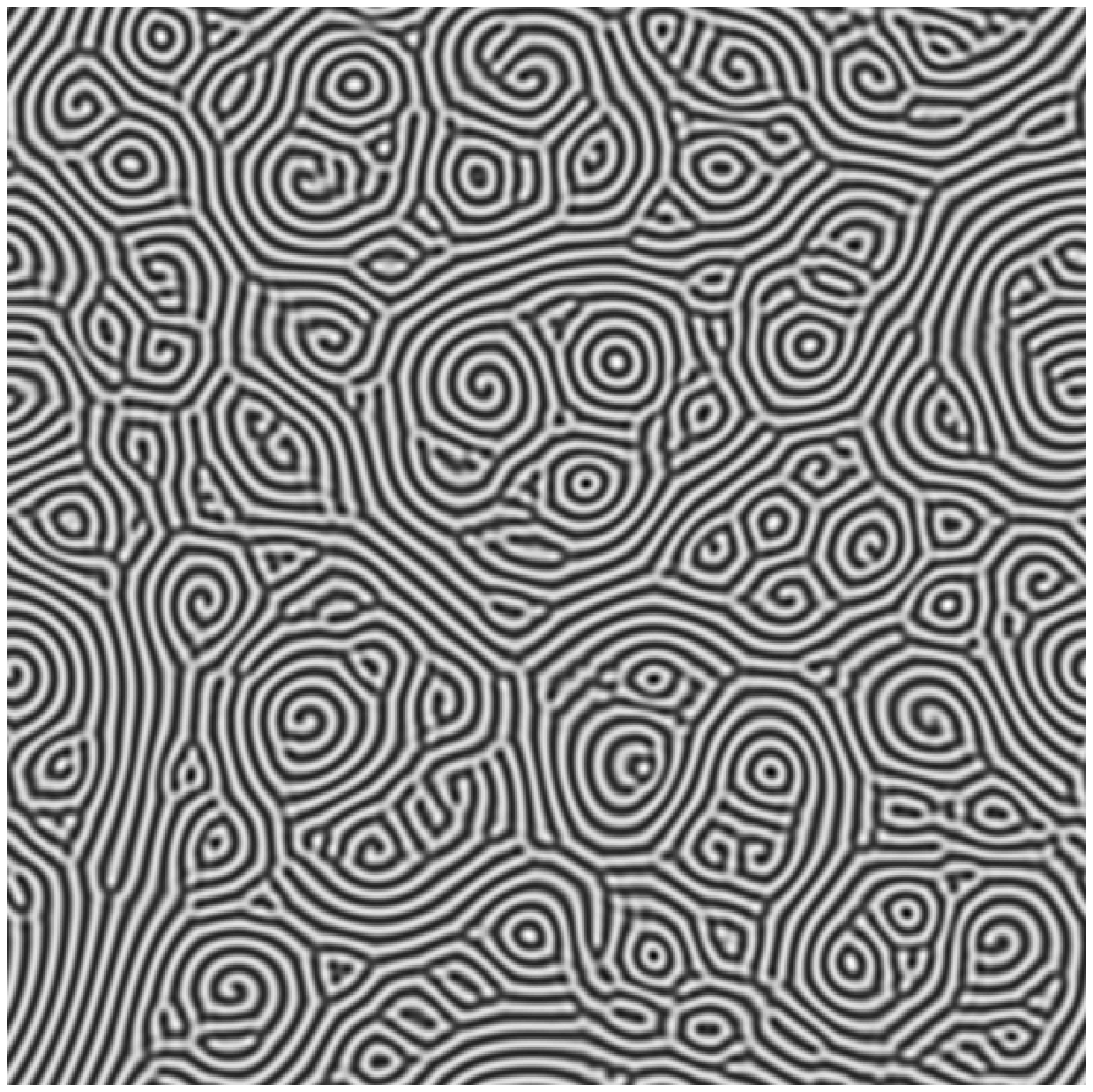} \\
	$c^2=0.1$ &  $c^2=0.5$ \vspace{0.4cm} \\
	\includegraphics[width=1.5in]{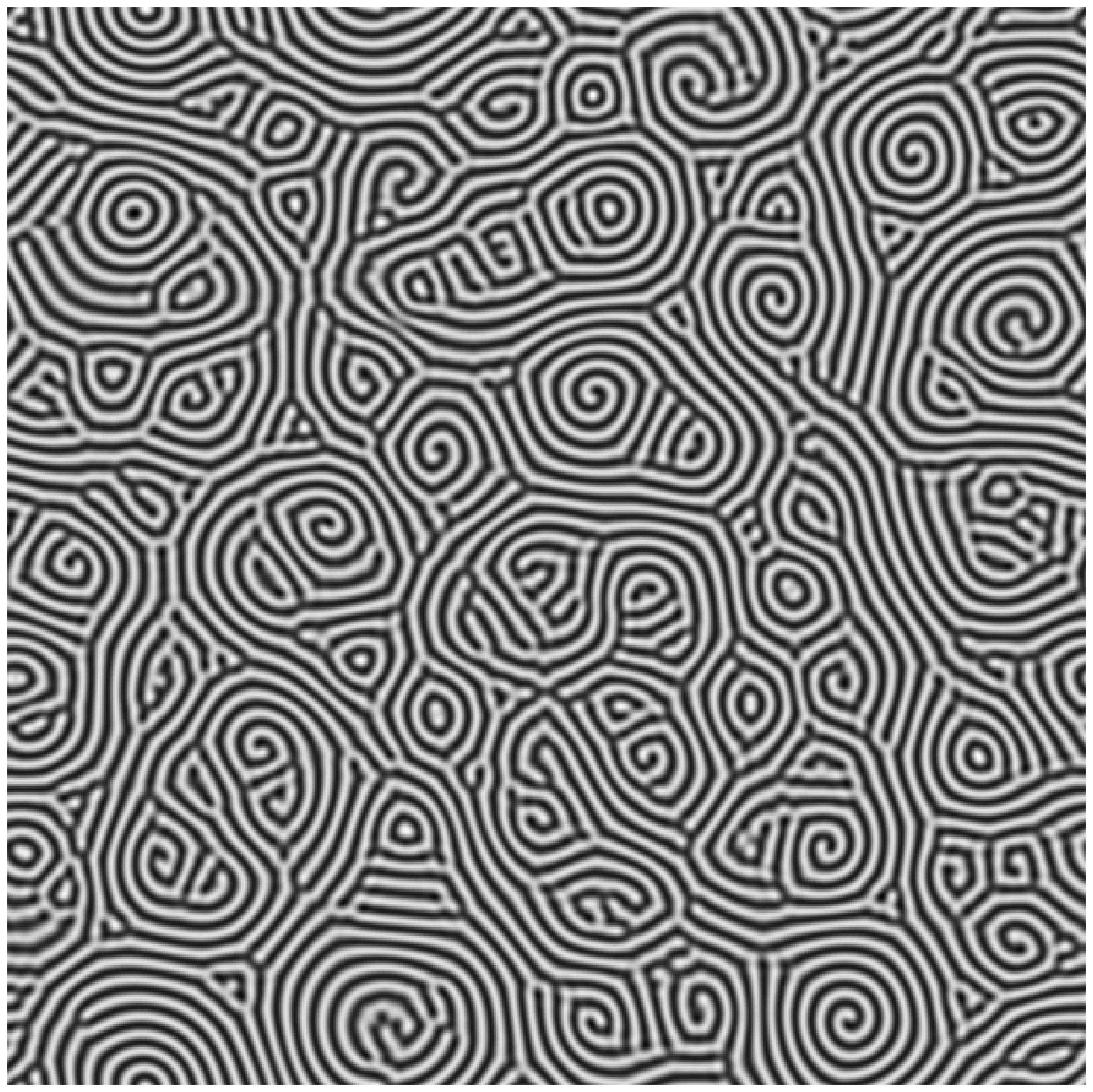}  \hspace{0.5cm} &
	\includegraphics[width=1.5in]{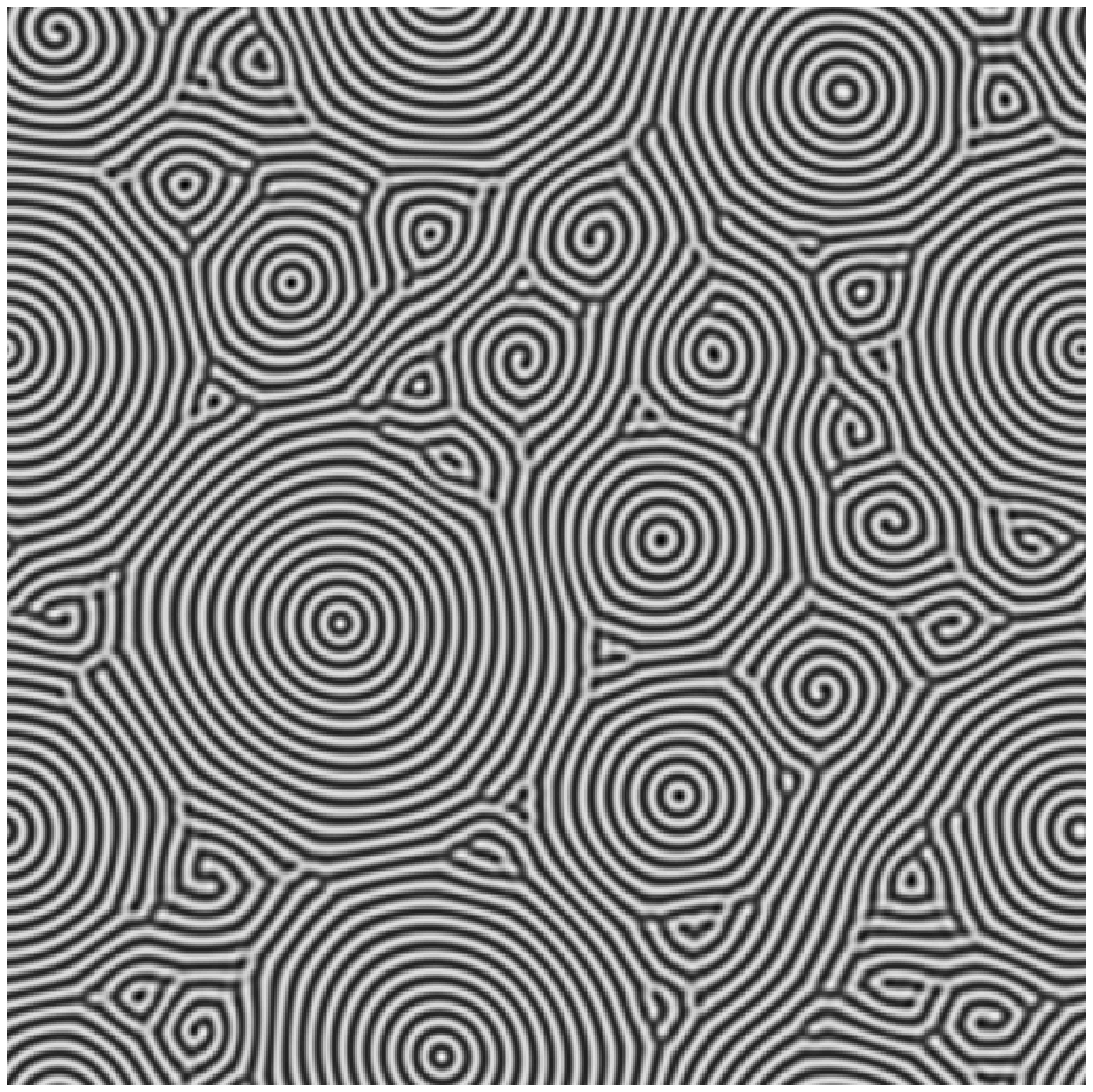} \\
	 $c^2=0.7$ &  $c^2=2$ \\  
\end{tabular}
\caption{The long-time patterns for Manneville's model over a range of mean 
flow strengths as determined by the value of $c^2$. The simulations
are initiated with random initial conditions and are integrated until
$t=10^6$. Other parameters are the same as Fig. \ref{fig:coarsening}.} 
\label{fig:patterns}
\end{figure}

Figure~\ref{fig:coarsening} illustrates the pattern evolution for $c^2=2$ using 
Manneville's model. The value of $c^2=2$ is the typical value used in the literature 
and corresponds to a weak mean flow.  At small time $t = 10^3$ the 
pattern is quite complex and contains many dynamic spiral and defect structures, 
similar to the scenario given in the simulations of Xi~\textit{et al.}~\cite{xi:1993}. 
The coarsening to target structures is evidenced by the pattern at
$t = 5 \times 10^4$ and represents the approximate duration of the
simulations by Schmitz~\textit{et al.} using the GSH equation~\cite{schmitz:2002}. 
This coarsening process is extremely slow as can be seen by comparing the patterns 
at longer times in Fig. \ref{fig:coarsening} (the pattern at $t=10^6$
for these simulation parameters is shown in the bottom right panel of
Fig.~\ref{fig:patterns}). Note that during the time evolution some big
targets can break up and form small defects which will then interact
and recombine with other targets or spirals, resulting in the process
of coarsening.

We have explored the long-time dynamics of the patterns by varying the strength 
of the mean flow over the parameter range of $0.1 \le c^2 \le 4$.  
Figure~\ref{fig:patterns} illustrates the patterns at $t=10^6$ for four different values 
of $c^2$. For the cases shown, the patterns resemble the state of spiral defect chaos for 
$c^2 \lesssim 0.7$, showing spatially complex structures with rapid
dynamics of small-scale spirals and localized defects. Although during
the evolution process some larger spirals or targets may be formed,
they are transients and will soon break up, with new small spiral
defects recreated; this procedure will repeat, but no long-lasting,
coarsened big targets or spirals can exist. However, for
larger $c^2$ (e.g., $=2$) the pattern has coarsened to a state
dominated by slowly moving target defects, as the case given in
Fig. \ref{fig:coarsening}. These observations indicate that strong
enough mean flows, as achieved via controlling parameter $c^2$ in the
model equations, are needed to reach a persistent chaotic or dynamic state, not 
only for breaking up large targets or spirals but also for locally recreating 
new small spiral-type defects to prevent the coarsening procedure.

\begin{figure}[htb]
\begin{tabular}{c}
	   \includegraphics[width=3.0in]{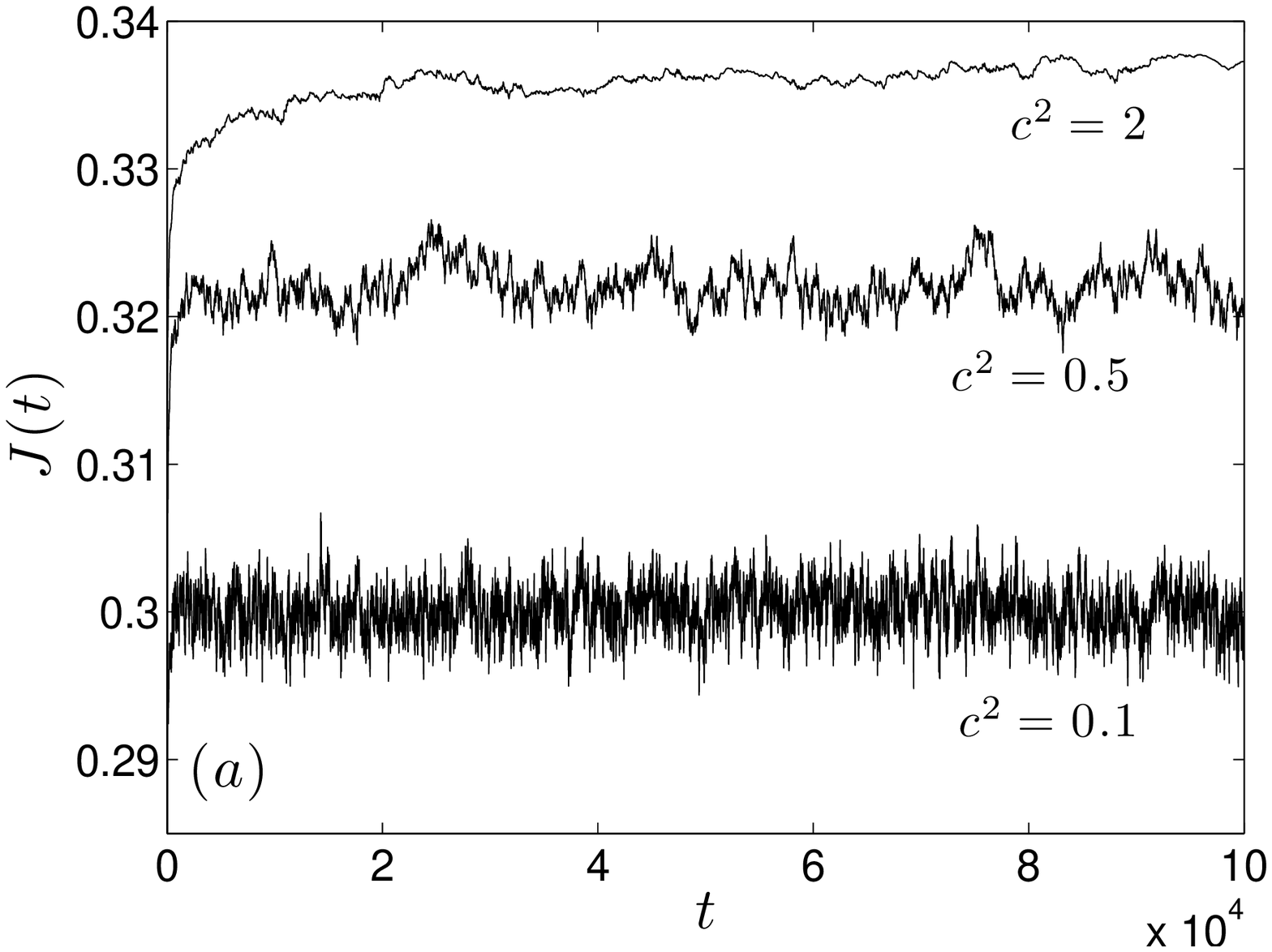} \\
	   \includegraphics[width=3.0in]{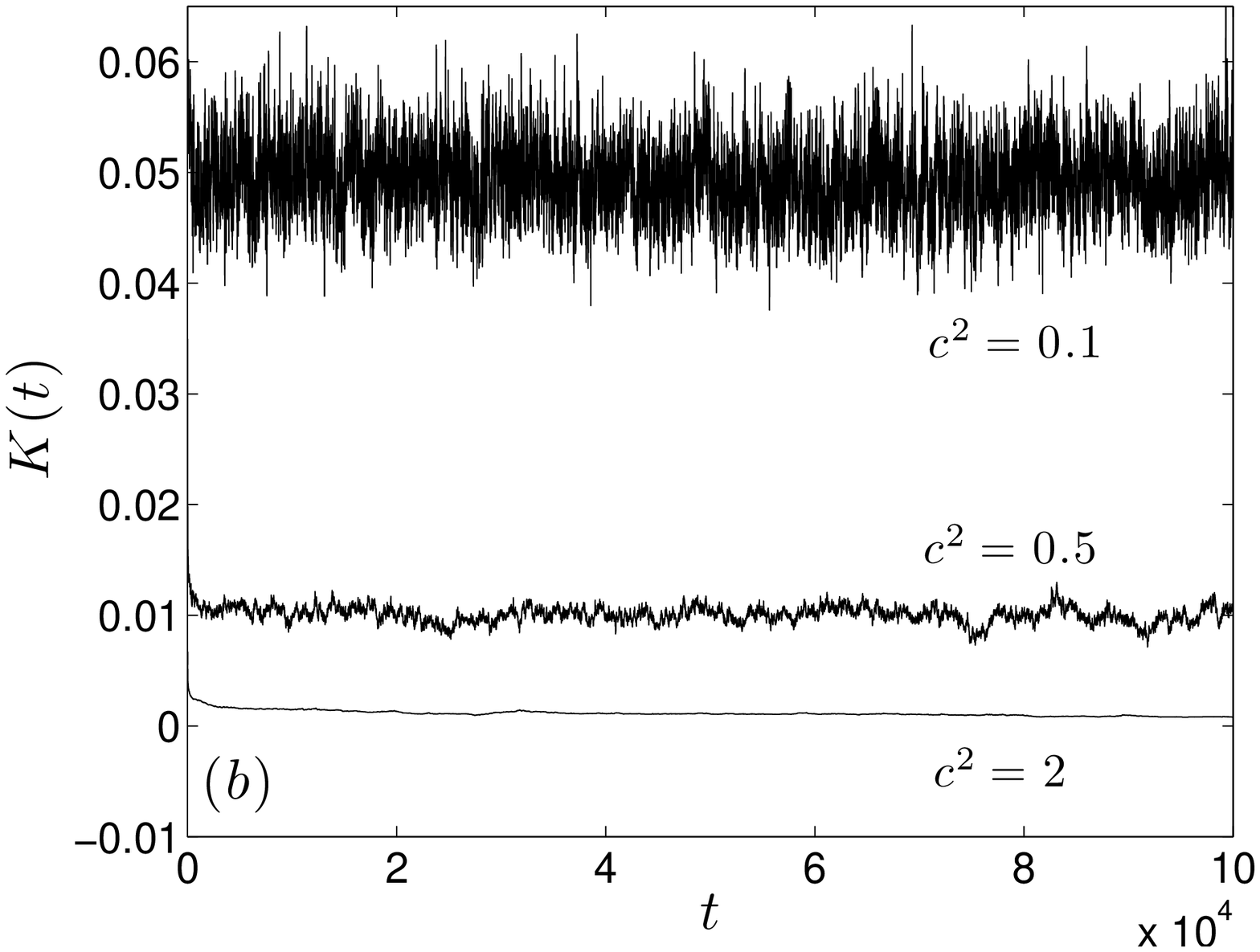}
\end{tabular} 
\caption{Time variation of the pattern dynamics in Manneville's model
for different magnitudes of the mean flow. The parameters used are the
same as those in Fig. \ref{fig:patterns}. (a)~Time variation of the
convective heat flux $J(t)$ given by Eq.~(\ref{eq:heat-flux}); 
(b)~Time variation of the average kinetic energy $K(t)$ defined by
Eq.~(\ref{eq:kinetic-energy}).}
\label{fig:dynamics}
\end{figure}

To further characterize the properties of system dynamics,
in Fig.~\ref{fig:dynamics} we examine the time variation of two global
quantities: the average kinetic energy $K(t)$ as defined in 
Eq.~(\ref{eq:kinetic-energy}), and the convective heat flux $J(t)$
which is given by
\begin {equation}
	J(t) = \left< \psi^2 \right>.
	\label{eq:heat-flux}
\end{equation}
Fig.~\ref{fig:dynamics}(a) indicates that as the strength of the mean
flow increases (i.e., the values of $c^2$ decrease), the magnitude of 
the heat flux decreases with increasing fluctuations about the mean 
value. Such fluctuations are a result of the pattern dynamics. More 
specifically, the dynamics of the defects 
yield the excursions to larger and smaller values of the heat flux. For 
the average kinetic energy as shown in Fig.~\ref{fig:dynamics}(b), 
both the magnitude of $K(t)$ and its fluctuations increase 
with increasing strength of the mean flow. Overall, Fig.~\ref{fig:dynamics} 
can be used to shed some qualitative insight upon the complexity of the 
dynamics. It is clear that for $c^2=2$ (which is the parameter used in
most previous studies) the disorder in the kinetic energy and heat flux is 
significantly reduced.

\begin{figure}[htb]
\begin{center}
	\includegraphics[width=3in]{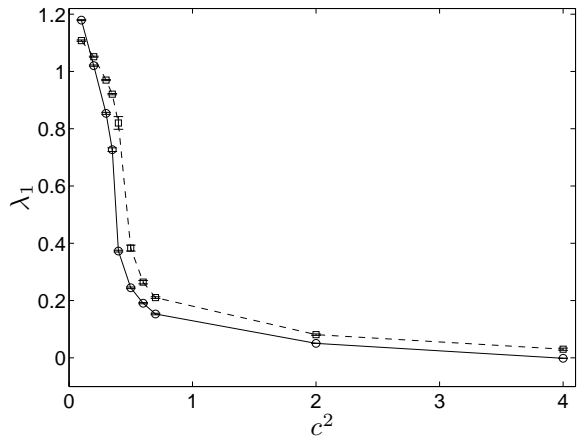}  
\end{center}
\caption{The variation of the leading order Lyapunov exponent $\lambda_1$ with 
the magnitude of the mean flow. The value of $\lambda_1$ is computed 
using results for the final $10^5$ time units. The 
circle symbols and the solid line are for Manneville's model using the system 
parameters $\epsilon=0.7$, $\sigma=2$, and $g_m=50$. The square symbols and the 
dashed line are for the GSH equation using the typical 
system parameters of most previous studies:
$\epsilon=0.7$, $\sigma=1$, and $g_m=50$.  The lines are only meant to guide the 
eye.  The error bars represent the standard deviation of 
$\lambda_1$ about its mean value for each value of $c^2$. The maximum value 
of the error bar for Manneville's model is $\sim10^{-3}$ and for the 
GSH equation is $\sim10^{-2}$.} 
\label{fig:lyapunov}
\end{figure}

Although global diagnostics such as the kinetic energy and heat flux provide 
qualitative insights into the complexity of system dynamics, they can
not be used to determine if the dynamics are chaotic.  In order to determine 
if the dynamics are chaotic we have also calculated the finite-time 
leading order Lyapunov exponent using the approach discussed in 
Sec.~\ref{section:simulations}. In Fig.~\ref{fig:lyapunov} we show the variation of 
$\lambda_1$ with the magnitude of the mean flow, for which $\lambda_1 > 0$ 
indicates chaos.  The circle symbols are the results for Manneville's model 
and the solid line is included only to guide the eye.  We emphasize
that these results are obtained from very long-time simulations to
ensure that any slow coarsening dynamics have been captured. The 
reported values of $\lambda_1$ are computed using 
the final $10^5$ time units.  The standard deviation of the fluctuations of $\lambda_1$ 
about its mean value is included as error bars. The maximum value of the error bar for the parameters 
explored is $\sim10^{-3}$.  We have also performed tests upon the data 
where $\lambda_1$ is computed over successive windows of time of 
various durations to quantify the presence of any slow trends. Our 
reported values of $\lambda_1$ appear to be the converged result 
and for the duration of the simulations explored we did not find any 
indication of slow trends. However, this does not rule out the possibility 
of even slower dynamics not captured in our simulations.

Our simulations indicate that the dynamics are weakly chaotic for $c^2 \approx 2$ 
where $\lambda_1 \approx 0.05$. This weak chaos is due to the slow 
dynamics of the pattern, as can be seen in Fig.~\ref{fig:patterns} where 
the targets are slowly moving among the more rapid dynamics of defects 
such as spirals, dislocations, etc. The value of $\lambda_1$ increases with 
the decreasing value of $c^2$ and thus the increasing magnitude of the 
mean flow as given in Fig.~\ref{fig:lyapunov}.

We have also included in Fig.~\ref{fig:lyapunov} the numerical results of the 
GSH equation to enable a more direct comparison with prior results in the 
literature. These results are shown as the square symbols and the dashed 
line. We have allowed $c^2$ to vary while the remaining
parameters are the typical values used in the previous studies:
$\epsilon=0.7$, $\sigma=1$, and $g_m=50$.  Overall our results indicate 
a very similar trend: the dynamics are weakly chaotic for $c^2\approx2$ 
and become increasingly chaotic for smaller $c^2$ and larger magnitudes 
of the mean flow.  We have performed the same series of tests as for 
Manneville's model to account for the presence of any slow coarsening process. The 
maximum magnitude of $\lambda_1$ error bars for the GSH equation is $\sim10^{-2}$.  

\begin{figure}
\begin{center}
	   \includegraphics[width=3in]{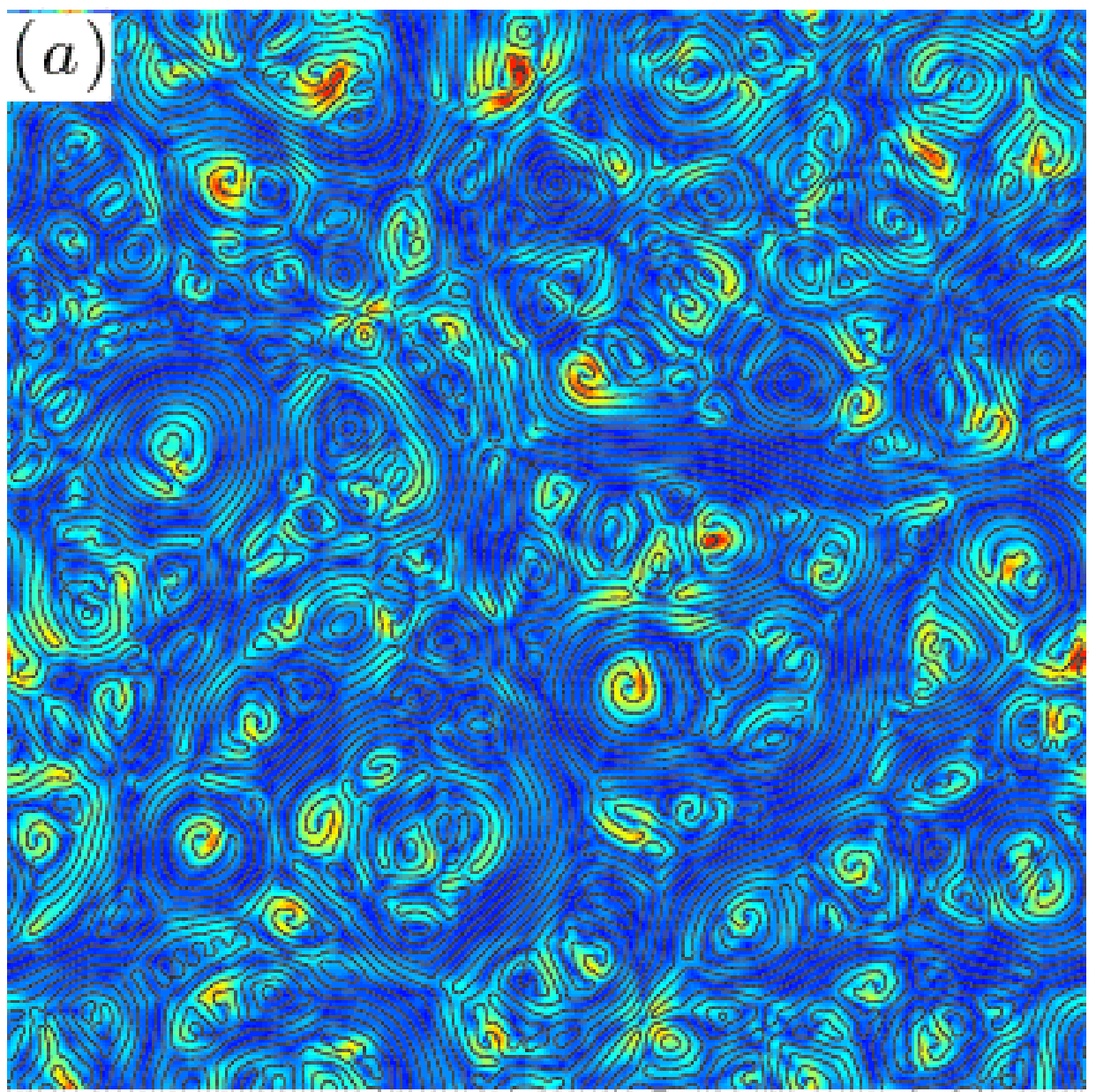}  \\
  	   \includegraphics[width=3in]{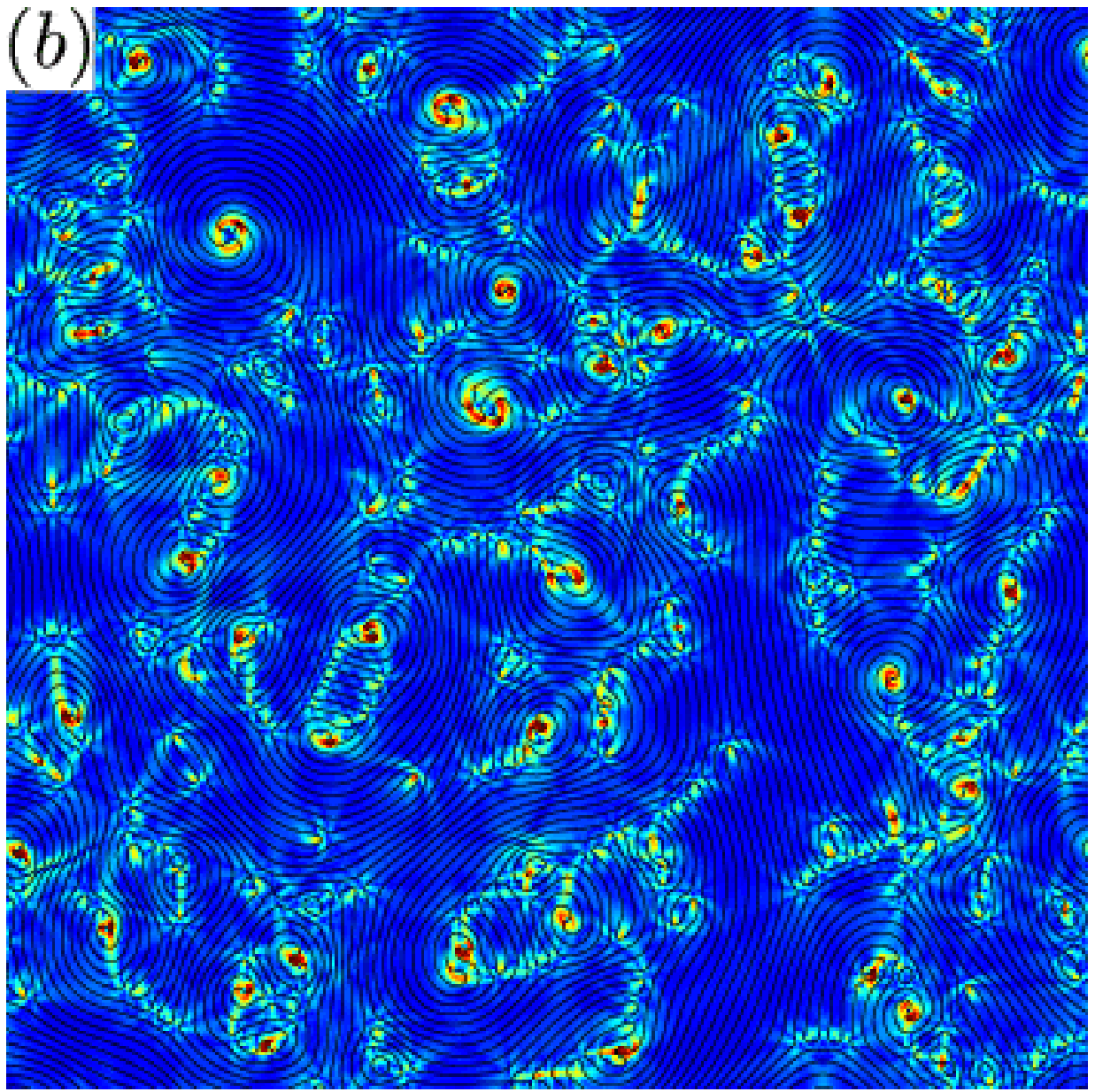}
\end{center}
\caption{(color) Spatial variation of the mean flow field for chaotic patterns containing 
many spiral defects.  The color contours represent the magnitude of the mean flow 
field, with red regions corresponding to large mean flow magnitude and blue regions 
corresponding to small mean flow magnitude. The black contours indicate convective roll 
boundaries. (a) Results for Manneville's equations at $t=10^6$ using 
system parameters ($\epsilon=0.7$, $\sigma=2$, $g_m=50$, $c^2=0.1$). (b) Results 
for the Boussinesq equations using system parameters
$\epsilon=0.7$, Pr$=1$, and $\Gamma=128$ at time $t=232.9$.} 
\label{fig:mean-flow}
\end{figure}

Although the connection between the Swift-Hohenberg equations and the Boussinesq 
equations of Rayleigh-B\'{e}nard is phenomenological, a direct comparison between 
the two can provide further physical insights. In this comparison we mainly focus 
on spatial features of the mean flow field for chaotic patterns containing 
a large number of spiral defects. In Fig.~\ref{fig:mean-flow}(a) we illustrate the relationship 
between the pattern and the magnitude of the mean flow. The image shown 
is for Manneville's model at $t=10^6$ with $c^2=0.1$.  The dynamics are chaotic with 
$\lambda_1 \approx 1.2$ as given in Fig.~\ref{fig:lyapunov}.  The color 
contours represent the magnitude of the mean flow field 
$|\mathbf{U}|$ where red indicates regions of large mean flow and blue 
indicates regions of small mean flow. The roll pattern is shown by the 
black lines which are given by contours of $\psi=0$. The mean flow tends 
to reach its local maximum at locations that contain defect structures
and remains large on a length scale of several roll wavelengths around the defect.

We also performed numerical simulations of the three-dimensional 
time-dependent Boussinesq equations that describe Rayleigh-B\'enard 
convection.  We used a parallel spectral approach that is discussed in 
detail elsewhere (c.f.~\cite{paul:2003}). We have chosen the reduced 
Rayleigh number $\epsilon=0.7$, the Prandtl number Pr=1,
and an aspect ratio of $\Gamma = 128$ for the spatial domain. The 
top and bottom boundaries have the no-slip fluid boundary 
condition and are held at constant temperature, while periodic 
boundary conditions are used on the sidewalls. The spatial variation of the 
mean flow is shown in Fig.~\ref{fig:mean-flow}(b). The mean flow is 
computed as the vertical average of the horizontal components of the 
fluid flow field, with red regions indicating large mean flow and blue 
regions corresponding to small mean flow. The black lines indicate the 
contours of the convection rolls.  Due to the large aspect ratio of the 
domain the computational cost of the numerical 
simulation is considerable, and the image shown is for a time $t=232.9$ 
where $t$ has been nondimensionalized in the usual manner using 
the time for heat to diffuse across the depth of the layer. Qualitatively,  
the convective roll pattern is quite similar to what is shown for the 
Swift-Hohenberg-type equations.  The spatial variation of the mean flow 
field, on the other hand, is different in terms of its rate of decay with 
the distance from a defect core.  The mean flow field is largest at the core regions 
of the defect structures and decays rapidly with distance away from the 
defect core. In most cases the length scale of this decay is approximately 
that of a single roll (see Fig.~\ref{fig:mean-flow}(b)), as compared to several 
rolls for results of Manneville's model shown in Fig.~\ref{fig:mean-flow}(a).
 
\begin{figure}
\begin{center}
	   \includegraphics[width=3.0in]{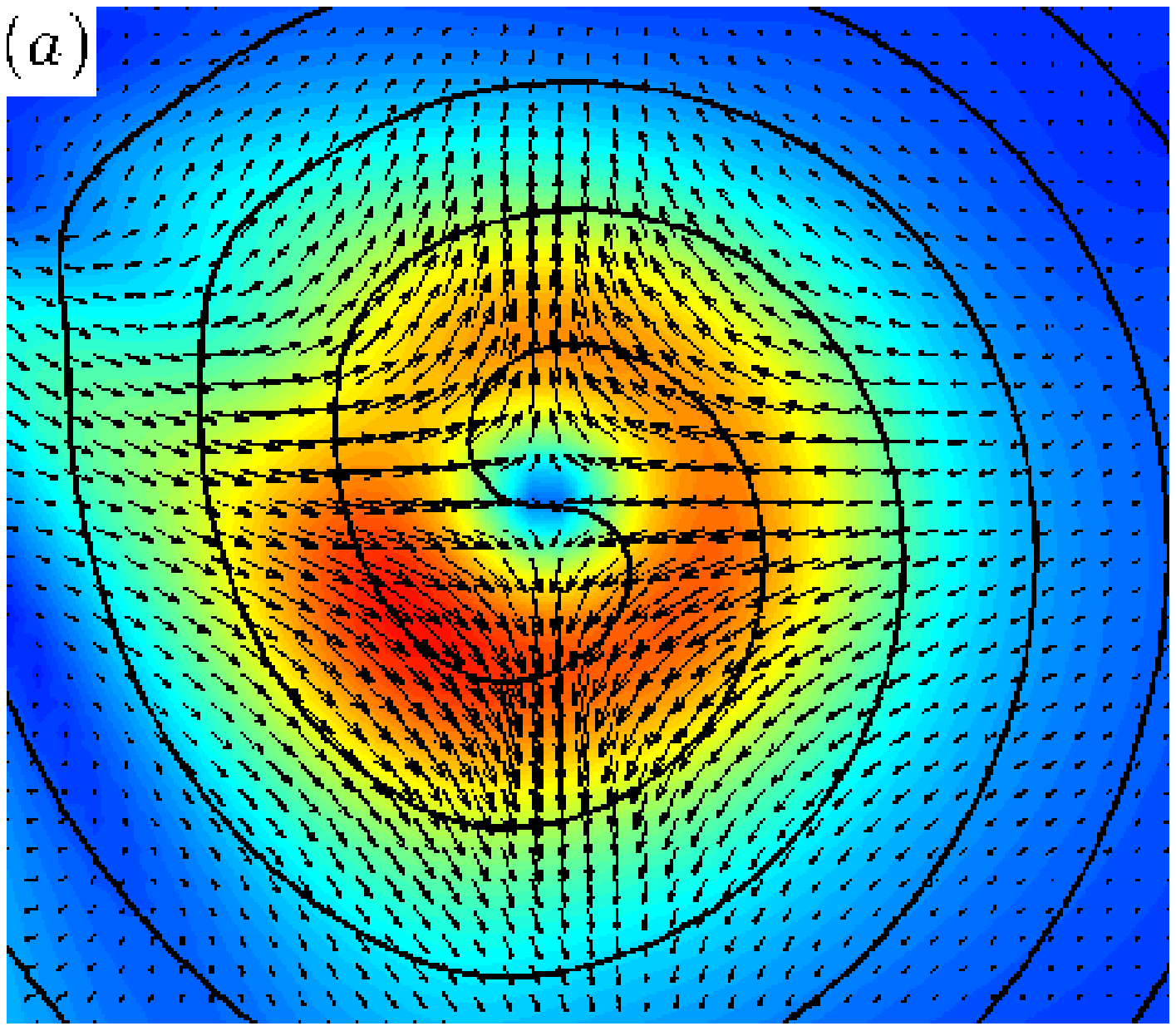}  \\
	   \includegraphics[width=3.0in]{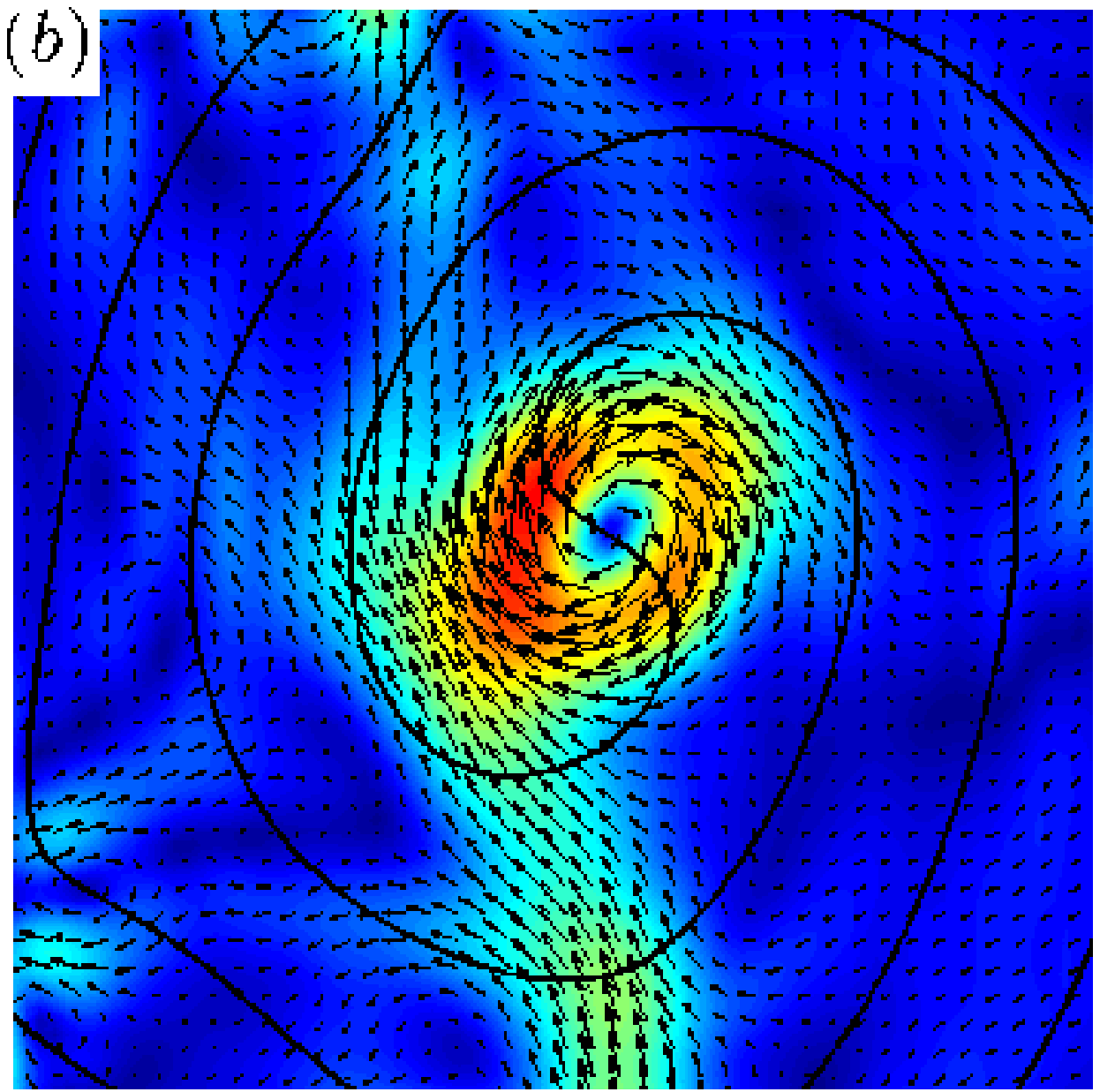}
\end{center}
\caption{(color) A close-up view of the spatial variation of the mean flow near a 
spiral defect in a chaotic pattern.  The color contours represent the magnitude of 
the mean flow as in Fig. \ref{fig:mean-flow}, the 
black contours indicate the convective roll boundaries, and the arrows are vectors for 
the mean flow. (a) Results using Manneville's equations with system parameters 
$\epsilon=0.7, \sigma=2$, $g_m=50$, and $c^2=0.4$ at time $t=7\times10^5$. 
(b)~Results from the Boussinesq equations using system parameters $\epsilon=0.7$, 
$\sigma=1$, and $\Gamma=128$ at time $t=232.9$.} 
\label{fig:spiral}
\end{figure}

The derivation of the Swift-Hohenberg equations is based upon a long wavelength 
approximation and hence the corresponding results are not expected to be accurate 
in the core regions of defect structures~\cite{swift:1977}.  This is illustrated 
in Fig.~\ref{fig:spiral} which shows a close-up view 
of the mean flow structure near a spiral defect, with panel (a)
for results of Manneville's model and panel (b) for the Boussinesq equations.  In 
both cases the spiral is rotating in a counter-clockwise direction. For the 
Swift-Hohenberg equations the mean flow exhibits a quadrupole structure centered upon the 
spiral. The magnitude of the quadrupole is spatially asymmetric and varies with the 
defect dynamics.  The mean flow for the Boussinesq equations is a vortex rotating 
in the opposite direction to the rotation of the spiral.  Our results 
for the Boussinesq equations are in agreement with those presented by 
Bodenschatz~\textit{et. al} (see Fig.~18 in Ref.~\cite{bodenschatz:2000}).

\section{Conclusion}
\label{section:conclusion}

Using generalizations of the Swift-Hohenberg equation we have explored the 
spiral defect chaos state and the role of the mean flow.  Our results show 
that it is possible to generate chaotic dynamics using Swift-Hohenberg-type 
model equations that resemble the spiral defect chaos of Rayleigh-B\'enard 
convection.  The important insight is that the strength of the mean flow must 
be large enough.  Reasonable parameter values of $\epsilon$, $\sigma$, and $g_m$ 
that lead to spiral defect chaos have been explored at some length in previous
literature.  However, the precise value of parameter $c^2$ depends on the approximations 
used in the derivation of Manneville's equations and the appearance of $c^2$ is 
phenomenological in writing down the GSH equations. Our results show that 
the dynamics vary strongly with the magnitude of $c^2$ and the commonly used 
value of $c^2=2$ yields a mean flow that is not strong enough to generate 
persistent dynamics that resemble spiral defect chaos. By choosing a smaller 
value of $c^2$ the dynamics are chaotic for as long as we have simulated ($t=10^6$ time units). 
Although we focused our discussion upon results generated using Manneville's equations,
our conclusions and insights also apply to the GSH equations. The 
particular choice of the form of the nonlinearity does not strongly affect the 
results in any significant way that we have found.  The necessity of a strong 
enough mean flow to support spiral defect chaos is in agreement with what 
has been found from the Boussinesq equations~\cite{chiam:2003}.

However, there are significant differences between numerical results from 
the Swift-Hohenberg-type model equations and those generated using the full three-dimensional 
Boussinesq equations.  The Swift-Hohenberg equations are not expected to 
capture the small scale features correctly due to the long-wavelength approximation 
involved, and we have quantified some aspects 
of this by comparing the mean flow fields around a single spiral defect.

We anticipate that our results will have several uses. First is that 
Swift-Hohenberg-type equations can be used as a model to study 
spatiotemporal chaos in a system with direct relevance to fluid phenomena such 
as Rayleigh-B\'enard convection.  This is a significant advantage since numerical 
simulations of the Boussinesq equations are computationally very expensive~\cite{paul:2003}.  
The Swift-Hohenberg model equations could be used to explore fundamental ideas of 
spatiotemporal chaos that are currently inaccessible to the full fluid equations. 
Examples include a detailed study of microextensivity~\cite{tajima:2002,karimi:2010} or the 
computation of the characteristic Lyapunov exponents~\cite{ginelli:2007,pazo:2008}. 
In addition, our comparison of the spatial variation of the mean flow around  
spiral defect core structures could be used to guide the development of more accurate 
theoretical descriptions of spiral defect chaos.

\bigskip

\noindent Acknowledgments: MRP and AK acknowledge support from NSF grant 
no. CBET-0747727. Z-FH acknowledges support from NSF under Grant No. DMR-0845264. 

\bibstyle{prsty}

\end{document}